\documentclass[showpacs,prb,preprintnumbers,amsmath,amssymb,twocolumn]{revtex4-1}
\usepackage[latin1]{inputenc}
\usepackage{graphicx}
\usepackage{dcolumn}
\usepackage{bm}
\usepackage[usenames]{xcolor}

\begin{document}

\title{Calculation of the magnetotransport for a spin-density-wave quantum critical theory in the presence of weak disorder}
\author{Hermann Freire$^{1}$}
\email{hermann\_freire@ufg.br}
\affiliation{$^{1}$ Instituto de Física, Universidade Federal de Goiás, 74.001-970, Goiânia-GO, Brazil}

\date{\today}

\begin{abstract}
We compute the Hall angle and the magnetoresistance of the spin-fermion model, which is a successful phenomenological theory to describe the physics of the cuprates and iron-based superconductors within a wide range of doping regimes. We investigate both the role of the spin-fermion interaction that couples the { large-momentum} antiferromagnetic fluctuations to the so-called ``hot-spots'' at the Fermi surface and also of an effective higher-order composite operator in the theory. { The latter operator} provides a scattering mechanism { such that} the momentum transfer for the fermions close to the Fermi surface can be small. We also include weak disorder that couples to both the bosonic order-parameter field and the fermionic degrees of freedom. { Since the quasiparticle excitations} were shown in recent works to be destroyed at the ``hot-spots'' in the low-energy limit of the model, we employ the Mori-Zwanzig memory-matrix approach that permits the evaluation of all transport coefficients without assuming well-defined Landau quasiparticles in the system. We then apply this transport theory to discuss universal metallic-state properties as a function of temperature and magnetic field of the cuprates from the perspective of their fermiology, which turn out to be in qualitative agreement with key experiments in those materials.
\end{abstract}
 

\maketitle

\newpage

\textbf{Introduction.} -- Transport properties in quantum critical matter \cite{Sachdev} continue to be one of the most formidable problems in strongly correlated systems and, despite many decades now
of intensive scrutiny, is still not entirely understood. One of the reasons for this conundrum is that the celebrated quasiparticle description in Landau's Fermi liquid theory and
the subsequent use of quantum Boltzmann's equation
are not expected to be adequate frameworks for the proper description of such systems. In this respect, probably the most famous example of a non-Fermi liquid is the strange metal phase \cite{Ong,Tyler}, 
which is observed in the cuprate superconductors, but there are many other examples (see, e.g., \cite{Matsuda,Analytis}). Even though such a phase is well-known to exhibit a non-Fermi-liquid dc resistivity that is linear in temperature ($T$),
the Hall angle $\theta_H$ is given by $\tan \theta_H\sim 1/T^2$, both inside the pseudogap phase and at optimal doping \cite{Ong,Ando,Raffy,Hussey}.
The latter observation indicates, somewhat surprisingly, a conventional scattering rate, which points to a Fermi-liquid-like result. This so-called separation of lifetimes
has remained a profound mystery up to the present day, which may suggest the application of alternative transport methods that do not necessarily rely on the concept of well-defined quasiparticles at low energies.

Materials that display spin-density-wave (SDW) quantum criticality \cite{Hertz,Chubukov1} abound in the literature. These may include the cuprates\cite{Daou}, the iron-based superconductors \cite{Matsuda,Analytis,Ni}, among many others (see, e.g., Ref. \cite{Raymond}). In the
particular case of the cuprates, it has been a long quandary whether its underlying physics is either driven by fermiology or is a result of strong interactions. {We point out that recent works have shown that in the case of a strongly underdoped regime, modifications of the so-called spin-fermion model that invoke topological order and an emergent gauge field are necessary to account for the non-double occupancy constraint, which should be enforced due to the strong interactions reflecting the influence of the Mott insulating phase at zero doping \cite{Ferraz,Berg}.
Despite this { cautionary} remark, we emphasize here that our starting point in this work will be a weak-to-moderate coupling approach to the problem (see, e.g., Ref. \cite{Freire4}), suitable for doping regimes not very far from optimal doping or in the overdoped regime.} In this respect, we note that a recent groundbreaking experimental 
work \cite{Hamidian} has provided evidence of a novel pair-density-wave in the cuprate superconductor Bi$_2$Sr$_2$CaCu$_2$O$_{8+x}$ using scanned Josephson tunneling microscopy, which was anticipated by theories that assume a crucial role of SDW criticality in those materials \cite{Wang2,Freire2}. 
In the SDW quantum critical scenario, the underlying mechanism for the formation of Cooper pairs in the superconducting phase of these compounds is rooted in the exchange of antiferromagnetic short-range SDW fluctuations \cite{Pines}, which are enhanced close
to the quantum critical point \cite{Chubukov1}. On the theoretical side, many works have described several aspects of the properties of the Abanov-Chubukov spin-fermion model using various complementary analytical \cite{Chubukov1,Metlitski,Efetov,Wang,SSLee,Freire2,Freire1,Ferraz,Tsvelik} and numerical \cite{Schattner} techniques. In this respect, it is now clear that the quasiparticle excitations are destroyed at the so-called hot spots (i.e., the intersection of the underlying Fermi surface with the antiferromagnetic zone boundary) at low energies \cite{Freire2,Schattner}. Lastly, we mention that the spin-fermion model is expected to provide a good description of many correlated systems both in the normal phase and in the symmetry-broken phases that appear at low temperatures, including of course the $d$-wave superconducting phase. 
 
In this work, we report a non-quasiparticle transport analysis based on the so-called Mori-Zwanzig memory matrix formalism \cite{Forster} of a weak disordered spin-fermion model with
the addition of an effective higher-order composite operator that couples the bosonic spin fluctuations to the entire Fermi surface of the system. We point out here that the inclusion of this composite interaction was first proposed in Ref. \cite{Hartnoll} and later demonstrated microscopically in Ref. \cite{Schmalian}. The present analysis will be performed in order to address the fundamental question, which revolves around the magnetotransport properties of this model due to the application of a magnetic field.
Naturally, one can expect some complications that could arise due to the application of a large field in those systems. It may cause, for instance, Fermi surface reconstruction \cite{Badoux}, in which the topology of Fermi surface in $\mathbf{k}$-space changes drastically as a result of the application of this field. This is indeed important to describe many correlated materials, including the cuprates. In the last years, there has been major progress experimentally in identifying novel electronic phases that emerge as low-energy instabilities in some cuprate superconductors due to the presence of large fields. In this respect, recently a unidirectional (nematic) long-range charge-density-wave order that shows up at low temperatures has been reported \cite{Chang} in the underdoped cuprate superconductor YBa$_2$Cu$_3$O$_{6+x}$ at moderate fields. However, one important point that we would like to stress is that we will focus here on the anomalous metallic phases that exist at intermediate temperatures in the model above which these symmetry-broken phases eventually intervene. As a result, we will leave the rather difficult analysis of the precise role of Fermi surface reconstruction in the present model at lower temperatures for a future work.  

This work is structured as follows. First, we define the effective spin-fermion model that we will analyze here with the inclusion of weak disorder. Then, we briefly review the memory matrix formalism in order to calculate the transport coefficients of this model. Next, we present our results concerning the magnetotransport properties (i.e., the Hall angle and the magnetoresistance) as a function of temperature and magnetic field. Finally, we conclude by offering a scenario that may explain the aforementioned separation of lifetimes that takes place in
the cuprate superconductors close to optimal doping, which is indeed verified experimentally.

\textbf{Spin-fermion model.} -- To begin with, we will depart from the spin-fermion model \cite{Chubukov1} with a higher-order effective composite interaction\cite{Hartnoll,Schmalian} included. The SDW quantum critical theory is therefore described by the following Lagrangian 

\vspace{-0.3cm}

\begin{eqnarray}\label{1}
\mathcal{L}&=&\sum_{\alpha}\bar{\psi}_{\alpha}(\partial_{\tau}+\bar{\varepsilon}_{\mathbf{k}})\psi_{\alpha}+\frac{1}{2}\chi^{-1}_{q}(\vec{\phi}\cdot\vec{\phi})+\frac{u}{4!}(\vec{\phi}\cdot\vec{\phi})^2\nonumber\\
&+&\lambda\sum_{\alpha\alpha'}\bar{\psi}_{\alpha}(\vec{\phi}\cdot\vec{\sigma}_{\alpha\alpha'}){\psi}_{\alpha'}+\lambda'\sum_{\alpha} \bar{\psi}_{\alpha}\psi_{\alpha}(\vec{\phi}\cdot\vec{\phi}),
\end{eqnarray}

\noindent where $\bar{\psi}_{\alpha}$ and ${\psi}_{\alpha}$ refer to the fermionic fields with spin projection $\alpha$, $\bar{\varepsilon}_{\mathbf{k}}={\varepsilon}_{\mathbf{k}}-\mu$ denotes the fermionic energy dispersion with a chemical potential $\mu$, $\vec{\phi}=(\phi_x,\phi_y, \phi_z)$ stands for the bosonic order-parameter field, $\vec{\sigma}=(\sigma_x,\sigma_y, \sigma_z)$ correspond conventionally to the Pauli matrices, and $\chi_{q}$ is the bosonic propagator. Lastly, $\lambda$ is the fermion-boson coupling constant and $\lambda'$ is the effective composite interaction constant.

Since the above model does not contain explicit umklapp processes, the dc conductivity, as it stands, is infinite at any temperature. For this reason, we must specify a mechanism that generates momentum relaxation in the model. At high temperatures, fermion scattering off phonons could of course provide such a mechanism. However, since we will be interested in the intermediate temperature regime of the present model, we will consider here impurity scattering as the main source of momentum relaxation. For simplicity, we will analyze here the effects of only two types of weak disorder in the system: one term that couples to the fermionic degrees of freedom in the model (i.e., a short-wavelength disorder) and another that couples to the bosonic order-parameter field (i.e., a long-wavelength disorder). As a result, we must add to Eq. (\ref{1}) the following expression

\vspace{-0.3cm}

\begin{eqnarray}\label{2}
\mathcal{L}_{imp}&=&\sum_{\sigma}V(\vec{r})\bar{\psi}_{\sigma}(\vec{r})\psi_{\sigma}(\vec{r})+\sum_{\sigma}m(\vec{r})\vec{\phi}(\vec{r})\cdot\vec{\phi}(\vec{r}),
\end{eqnarray}

\noindent which should conform to the standard Gaussian disorder averages: $\langle\langle V(\vec{r}) \rangle\rangle=\langle\langle m(\vec{r}) \rangle\rangle=0$,
$\langle\langle V(\vec{r})V(\vec{r'}) \rangle\rangle=V_0^2\delta^2(\vec{r}-\vec{r'})$, and $\langle\langle m(\vec{r})m(\vec{r'}) \rangle\rangle=m_0^2\delta^2(\vec{r}-\vec{r'})$, where $V_0$ is a random potential
for the fermionic field and the parameter $m_0$ is a random mass term. 

\textbf{Memory matrix calculation.} -- Since we will be primarily interested here in the effects of a magnetic field $B$ in the present SDW quantum critical theory, we will start by including its effects in the model from the outset. For simplicity, we will consider only perturbative effects of the field $B$. We will then use the so-called Mori-Zwanzig memory matrix formalism that does not assume the existence of low-lying quasiparticles in the system \cite{Forster,Rosch,Patel,Patel2,Hartnoll3,Lucas,Freire5,Hartnoll2}.

For $B\neq 0$, the matrix of generalized conductivities $\hat{\sigma}$ as a function of temperature, magnetic field and frequency $\omega$ is given by

\vspace{-0.3cm}

\begin{equation}\label{sigma}
\hat{\sigma}(\omega,T,B)=\frac{\hat{\chi}^R(T)}{(\hat{M}+\hat{N}-i\omega{\hat{\chi}^R(T)})[{\hat{\chi}^R}(T)]^{-1}},
\end{equation}

\noindent where $\hat{M}$ is the memory matrix and $\hat{N}$ is a time-reversal symmetry breaking matrix that arises due to the magnetic field (both to be defined below). The quantity $\chi^R_{\mathcal{A}\mathcal{B}}(\omega=0,T)$ is the matrix of the static retarded susceptibilities of some conserved (or almost conserved) operators in the model, denoted arbitrarily as operators
$\mathcal{A}$ and $\mathcal{B}$. This susceptibility is conventionally defined as 
$\chi_{\mathcal{A}\mathcal{B}}(i\omega,T)=\int_{0}^{1/T}d\tau e^{i\omega\tau}\langle T_{\tau}\mathcal{A}^{\dagger}(\tau)\mathcal{B}(0)\rangle$, where the retarded susceptibility is of course given by $\chi^R_{\mathcal{A}\mathcal{B}}(\omega)=\chi_{\mathcal{A}\mathcal{B}}(i\omega\rightarrow\omega +i0^{+})$. Besides, $\langle ...\rangle$ refers to the
grand-canonical ensemble average, $T_{\tau}$ corresponds to the time-ordering operator, and the volume $V$ of the system, for simplicity, has been set equal to unity. 

The memory matrix $\hat{M}(T)$ can be computed from the following exact expression

\vspace{-0.3cm}

\begin{equation}\label{4}
\hat{M}_{\mathcal{A}\mathcal{B}}(T)=\int_{0}^{1/T}d\tau\left\langle \dot{\mathcal{A}}^{\dagger}(0)Q\frac{i}{\omega-Q\hat{L}Q}Q\dot{\mathcal{B}}(i\tau)\right\rangle,
\end{equation}

\noindent where $\hat{L}$ stands for the Liouville operator, which is defined as $\hat{L}\mathcal{A}=[H,\mathcal{A}]=-i\dot{\mathcal{A}}$, with $H$ corresponding to the Hamiltonian of the system and $Q$ is a projection operator that projects out of a operator space
spanned by the conserved (or nearly conserved) operators represented generically by $\{\mathcal{A},\mathcal{B},...\}$. As will become clear shortly, the memory matrix will quantify the relaxation mechanism of all the conserved (and almost conserved) operators in the present transport theory. As for the time-reversal noninvariant matrix $\hat{N}$, its matrix elements are in turn given by: $N_{P_i P_j}=\chi_{P_i\dot{P}_j}$. Therefore, for $B=0$, one can demonstrate straightforwardly that the matrix $\hat{N}$ vanishes identically for any temperature, whereas for $B\neq 0$ this matrix becomes finite instead.

Using Noether's theorem, and in view of the fact that the Lagrangian in Eq. (\ref{1}) is invariant under space translation and global $U(1)$ symmetry, the electrical current given by $\mathbf{J}=\int d^2 x \,\mathbf{j(x)}$, and the canonical momentum operator given by $\mathbf{P'}=\int d^2 x \,\mathbf{p(x)}$ are both conserved at the classical level. They read as follows
\vspace{-0.3cm}

\begin{eqnarray}\label{5}
\mathbf{J}&=&-\frac{i}{m}\sum_{\sigma}\int d^2 x \nabla \bar{\psi}_{\sigma}\psi_{\sigma},\\ \label{5.1}
\mathbf{P'}&=&\int d^2 x\left[i\sum_{\sigma}\nabla \bar{\psi}_{\sigma}\psi_{\sigma} + (\partial_t \phi) \nabla\phi \right].
\end{eqnarray}

\noindent Consequently, at the quantum level, these operators are expected to have the longest relaxation timescales in the system, and, for this reason, we will assume that they must play a central role in the transport properties of the present model. 

\begin{figure}[t]
\includegraphics[width=3.1in]{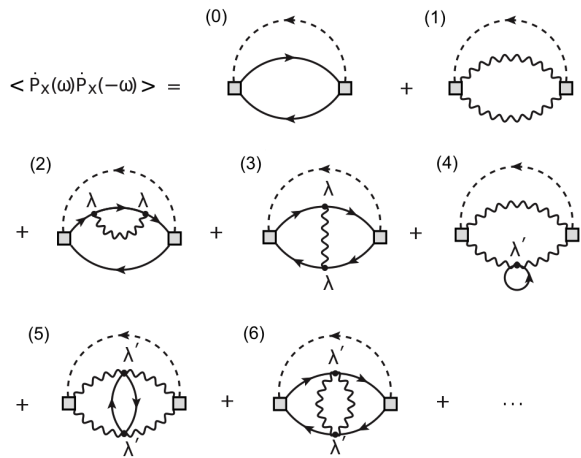}
\caption{Diagrams up to second order in the couplings related to the calculation of $G^{R}_{\dot{P_x}\dot{P_x}}(\omega,T)$. Solid lines denote the fermionic propagators, while wavy lines represent the bosonic propagators. The coupling $\lambda$ stands for the scattering between any pair of hot spots at the Fermi surface, whereas the coupling $\lambda'$ is the coupling related to the composite operator that couples to the whole Fermi surface of the model. The dotted lines refer to the impurity lines and carry only internal momentum and external bosonic energy $\omega$.}\label{Feynman_diagrams}
\end{figure}

One important point to emphasize is that, by adding weak disorder [Eq. (\ref{2})] to the Lagrangian [Eq. (\ref{1})], the canonical momentum [Eq. (\ref{5.1})] is not
conserved any longer, due to the breaking of the translation symmetry in the system. Therefore, we get the following equation of motion for the canonical momentum

\vspace{-0.3cm}

\begin{eqnarray}\label{6}
i\dot{\mathbf{P'}}&=&\int \frac{d^2\mathbf{q}}{(2\pi)^2}\int \frac{d^2\mathbf{k}}{(2\pi)^2}\mathbf{k}\bigg[V(\mathbf{k})\sum_{\sigma}\bar{\psi}_{\sigma}(\mathbf{k}+\mathbf{q})\psi_{\sigma}(\mathbf{k})\nonumber\\
&+&m(\mathbf{k})\phi(\mathbf{q})\phi(\mathbf{-q-k})\bigg].
\end{eqnarray}

\noindent Regarding the propagator for the fermions, we will focus here on the vicinity of the Fermi surface. For this reason, we approximate the fermionic energy dispersion as $\bar{\varepsilon}_{\mathbf{k}}=\vec{v}_{\mathbf{k}}.\mathbf{k}+u'\mathbf{k}^2$, where $\vec{v}_{\mathbf{k}}$ corresponds to the Fermi velocity of the system and $u'$ is naturally connected to the curvature of the Fermi surface.
Moreover, we will assume in what follows that the bosonic Green's function at intermediate temperatures is described by the dynamical critical exponent $z_b=1$ in the spin-fermion model, i.e.,

\vspace{-0.3cm}

\begin{eqnarray}\label{7}
\chi(q_0,\mathbf{q})&=&\frac{1}{q_{0}^{2}+\mathbf{q}^2+R(T)},
\end{eqnarray}

\noindent where $q_0$ stands for the bosonic Matsubara frequency, $\mathbf{q}$ is the bosonic momentum centered around the SDW ordering wavevector $(\pi,\pi)$ and $R(T)=4\ln^2[(\sqrt{5}+1)/2]T^2$ is
an infrared cutoff in the theory, which was computed previously in Ref. \cite{Chubukov3}. This
assumption can also be justified by invoking the RG program developed by Sur and Lee \cite{SSLee}, in which they managed to obtain a novel infrared-stable nontrivial fixed point in the present model for $d=3-\epsilon$ dimensions that is perturbatively controlled.

When an external magnetic field is applied to the system, the Lagrangian should be altered to $\mathcal{L}\rightarrow \mathcal{L}+\int d^2 x \,\mathbf{j}(\mathbf{x})\cdot\mathbf{A}(\mathbf{x})$, where $\mathbf{j}$ is the current density and $\mathbf{A}$ is the vector potential such that $\mathbf{B}=\nabla \times \mathbf{A}$. In this case, the physical momentum $\mathbf{P}$ becomes different from the canonical momentum $\mathbf{P'}$ through the relation

\vspace{-0.3cm}

\begin{eqnarray}
\mathbf{P}=\mathbf{P'}-\int d^2 x \, n(\mathbf{x}) \mathbf{A}(\mathbf{x}),
\end{eqnarray}

\noindent where $n$ is the charge density, which satisfies the continuity equation: $\partial n/\partial t + \nabla\cdot\mathbf{j}(\mathbf{x})=0$. Regarding the underlying symmetries of the model, the time-reversal ($\hat{\mathcal{T}}$) symmetry in the above case becomes broken due to the application of the magnetic field. Moreover, as will become clear soon, the role of the charge conjugation ($\hat{\mathcal{C}}$) symmetry in the model will also be important and will be discussed in more detail below. 

Using the following convenient gauge choices given by $\mathbf{A}=B(-y,0)$ and $\mathbf{A}=B(0,x)$, we obtain that $N_{P_x P_y}=-N_{P_y P_x}=-B\chi_{J_xP_x}$,
whereas $N_{P_x P_x}=N_{P_y P_y}=0$, i.e.,
\begin{equation}
\hat{N}=\left( \begin{array}{ccc}
0 & -B\chi_{J_xP_x} \\
B\chi_{J_xP_x} & 0 \end{array} \right).
\end{equation}

\noindent From Eq. (\ref{sigma}), the matrix of the dc conductivities will be naturally given by

\vspace{-0.3cm}

\begin{equation}
\hat{\sigma}_{dc}=\hat{\sigma}(\omega\rightarrow 0)=\hat{\chi}^R(T)(\hat{M}+\hat{N})^{-1}\hat{\chi}^R(T),
\end{equation}

\noindent where, to lowest order in $B$, we obtain

\vspace{-0.3cm}

\begin{equation}
(\hat{M}+\hat{N})^{-1}\approx \frac{1}{\det(\hat{M}+\hat{N})}\left( \begin{array}{ccc}
M_{P_yP_y} & B\chi_{J_xP_x} \\
-B\chi_{J_xP_x} & M_{P_xP_x} \end{array} \right).
\end{equation}

\noindent Thus, the expressions for the electrical conductivity $\sigma_{xx}$ and the Hall conductivity $\sigma_{xy}$ will finally be given by  

\vspace{-0.3cm}

\begin{eqnarray}
{\sigma}_{xx}(T,B)&=&\frac{\chi^2_{J_x P_x} M_{P_x P_x}}{(M_{P_x P_x}^2+B^2 \chi^2_{J_x P_x})},\\
{\sigma}_{xy}(T,B)&=&\frac{B\chi^3_{J_x P_x}}{(M_{P_x P_x}^2+B^2 \chi^2_{J_x P_x})}.
\end{eqnarray}

\noindent First, we proceed to analyze the static retarded susceptibility $\chi_{J_x P_x}(T)$ of the present system, which, to leading order, turns out to be equal to the noninteracting susceptibility, i.e.,

\vspace{-0.3cm}

\begin{eqnarray}\label{9}
\chi_{J_x P_x}(T)&=&\frac{1}{m}\int \frac{d^2\mathbf{k}}{(2\pi)^2} k_x^2 \frac{[n_F(\bar{\varepsilon}_{\mathbf{k}})-n_F(\bar{\varepsilon}_{\mathbf{k+q}})]}{(\bar{\varepsilon}_{\mathbf{k}}-\bar{\varepsilon}_{\mathbf{k+q}})}
\nonumber\\
&\approx&\frac{m}{\pi} \varepsilon_F+O(e^{-\beta \varepsilon_F}),
\end{eqnarray}

\noindent where $\varepsilon_F$ is the Fermi energy, $n_F(\varepsilon)=1/(e^{\beta\varepsilon}+1)$ is the Fermi-Dirac distribution, and $\beta=1/T$ is the inverse temperature.
Therefore, we can observe from the above result that the susceptibility $\chi_{J_x P_x}(T)$, in fact, does not possess any temperature dependence to leading order. In this way, the temperature dependence in both
the electrical conductivity and the Hall conductivity of the spin-fermion model
will originate from the memory matrix itself given by $M_{P_x P_x}(T)$.

The next step of our analysis consists of calculating perturbatively the memory matrix in the present system. For this reason, we will consider here that the parameters $\lambda$, $\lambda'$, $V_0$, and $m_0$ are effectively small. Therefore, in view of the fact that the Eq. (\ref{6}) is of order linear in $V_0$, and $m_0$, the
most important contribution to the memory matrix will turn out to be quadratic in these parameters. Moreover, reasoning in a similar way, the leading contribution to the Liouville operator will be given by the noninteracting value ($L\approx L_0$) and the most important contribution of the grand-canonical ensemble average should be expressed in terms of the noninteracting Hamiltonian of the system. As a consequence, it can be shown that the most relevant contribution to the memory matrix is given by

\vspace{-0.3cm}

\begin{eqnarray}\label{10}
M_{P_x P_x}(\omega\rightarrow 0,T)&=& \lim_{\omega\rightarrow 0}\frac{\text{Im}\,G^{R}_{\dot{P_x}\dot{P_x}}(\omega,T)}{\omega},
\end{eqnarray}

\noindent where $G^{R}_{\dot{P_x}\dot{P_x}}(\omega,T)=\langle\dot{P_x}(\omega)\dot{P_x}(-\omega)\rangle$ denotes the corresponding retarded Green's function at finite temperature. The Feynman diagrams related to this calculation are given explicitly in Fig. 1.

\textbf{Results.} -- From Fig. 1, one can see that our calculated memory matrix can be written schematically in a Matthiessen-rule-like way as

\vspace{-0.4cm}

\begin{eqnarray}\label{11}
M_{P_xP_x}(T)=\sum_{i=0}^{6}M^{(i)}(T)+\dots,
\end{eqnarray}

\noindent where the index $i$ refers, respectively, to each Feynman diagram shown in this figure. The first
Feynman diagram [i.e., with label (0)] represents the lowest order diagram related to the coupling of short-wavelength disorder to the 
fermions in the model. The computation of this diagram yields 

 \vspace{-0.3cm}

\begin{eqnarray}\label{12}
M^{(0)}=V_0^2\,\text{Im}\left\{\sum_{i,j,i\neq j}\frac{Q^{2}_{ij}}{\omega}\int_{\mathbf{k},\mathbf{q}}\frac{[n_F(\bar{\varepsilon}_{\mathbf{q}})-n_F(\bar{\varepsilon}_{\mathbf{k+q}})]}{\omega+\bar{\varepsilon}_{\mathbf{q}}-\bar{\varepsilon}_{\mathbf{k+q}}+i0^+}\right\},\nonumber\\
\end{eqnarray}

\noindent where the limit $\omega\rightarrow 0$ is implicit in the above expression and
$\int_{\mathbf{k}}=\int d^2 \mathbf{k}/(2\pi)^2$. By using the well-known identity $1/(x+i0^+)=\mathcal{P}(1/x)-i\pi\delta(x)$, we get the following result
$M^{(0)}=-\sum_{i,j,i\neq j}Q^2_{ij} V_0^2  \Lambda^2/(4\pi^3 |{\vec{v}_{i}}\times{\vec{v}_{j}}|)$, where $\Lambda$ is a ultraviolet cutoff that must be imposed in the integration over all the energies in the theory and the quantity $Q_{ij}$ is simply the (large) momentum transfer connecting the hot spots denoted by arbitrary indices $i,j$ in the theory. Of course, this turns out to be a temperature-independent contribution to the memory matrix. 

As for diagram (1) depicted in Fig. 1, this scattering process refers to
the coupling of the random mass term to the bosonic order-parameter field in the present model. Therefore, it evaluates to

\vspace{-0.3cm}

\begin{eqnarray}\label{13}
M^{(1)}(T)&=&\frac{m_0^2}{2}\,\text{Im}\Bigg\{\int_{\mathbf{k},\mathbf{q}}k_x^2\int_{E_1,E_2}\pi^2\text{sign}(E_1)\text{sign}(E_2)\nonumber\\
&\times&(E_1^2-R(T))\theta(E_1^2-R(T))\theta(E_2^2-R(T))\nonumber\\
&\times&\left(\frac{1}{\omega}\right)\frac{[n_B(E_2)-n_B(E_1)]}{\omega+E_2-E_1+i0^+}\Bigg\},
\end{eqnarray}

\noindent where $\theta(x)$ is the standard step function and the function $n_B(\varepsilon)=1/(e^{\beta\varepsilon}-1)$ denotes the Bose-Einstein distribution. Calculating the above expression, we obtain the following result

\vspace{-0.3cm}

\begin{eqnarray}\label{13a}
M^{(1)}(T)\approx \left(\frac{1.77 m_{0}^2}{16\pi^2}\right)T^2.
\end{eqnarray}

\noindent Interestingly, the above contribution to the memory matrix turns out to be Fermi-liquid-like. Another important property of the above expression is that the prefactor
turns out to be completely isotropic and doping-independent.

The diagram (2) that represents a scattering process involving a self-energy correction to the fermionic propagator in the model naturally evaluates to zero. In fact, it can be shown straightforwardly that all diagrams for the memory matrix with self-energy corrections to both fermionic and bosonic propagators [see also, e.g., diagram (4)] vanish in the present calculation. 

Diagram (3) in Fig. 1 refers to an Altshuler-Aronov-type correction. This particular Feynman diagram has been already discussed in the context of a two-band spin-fermion model inside the quantum critical phase in Ref. \cite{Patel}, and it is known to give a non-Fermi-liquid contribution to the memory matrix. Here, we obtain a similar result for a one-band model with a noninteracting Fermi surface appropriate for the description of the metallic phase of the cuprates around optimal doping, i.e.,

\begin{eqnarray}\label{16}
&&M^{(3)}(T)\approx \sum_{i,j,i\neq j}\sum_{\alpha,\beta}\left(\frac{0.001 V_{0}^2 \lambda^{2} Q^{2}_{ij}
}{ | \vec{v}_{i\alpha}\times \vec{v}_{i\beta}| | \vec{v}_{j\alpha}\times \vec{v}_{j\beta}|}\right)T.
\nonumber\\
\end{eqnarray}

\noindent The above term is related to the linear-in-$T$ resistivity of the strange metal phase that it is notorious to exist in the present model. This result is connected
to the contribution emerging from inter-hot-spot scattering processes on the Fermi surface. In other words, if there are either no hot spots in the theory or if these hot spots are gapped out (as can happen, e.g., both inside the pseudogap phase of the underdoped cuprates and also in the same materials at the extremely overdoped regime), this linear term will clearly not appear in the resistivity of the system. This observation is consistent with experimental data.

The calculation of the remaining diagrams (5) and (6) in Fig. 1 are tedious but their evaluation is relatively straightforward. It can be shown that the leading term associated to the diagram (5) evaluates to
$M^{(5)}(T)\sim m_0^2\lambda'^2 \Lambda^2$. Since this latter result is clearly temperature independent, it will contribute to a residual resistivity $\rho_0$ in the present system. For simplicity, we will omit this term  from this point on. As for the diagram (6), we obtain that its leading contribution is naturally given by $M^{(6)}(T)\sim V_0^2\lambda'^2 T^4$. This latter term turns out to be subleading to all the other contributions to the memory matrix calculated in this work. For this reason, although this additional temperature dependence to the dc resistivity of the model can be very interesting on its own right, it might be unfortunately very difficult to detect it from an experimental point of view. For this reason, we will overlook this contribution in the transport analysis that follows. 

We can start from the calculation of the effect of the magnetic field $B$ on the magnetoresistance inside the pseudogap phase of the cuprates, where it is well-known that the quasiparticle excitations at the hot spots are completely destroyed in the low-energy limit \cite{Freire10,Metlitski,Schattner}. In this particular case, the corresponding magnetoresistance should be given by

\vspace{-0.3cm}

\begin{eqnarray}
\frac{\Delta\rho_{xx}}{\rho_{xx}}=\frac{[\rho_{xx}(B)-\rho_{xx}(0)]}{\rho_{xx}(0)}\sim \frac{B^2}{T^4}.
\end{eqnarray}

\noindent This is in good agreement with some experiments in the underdoped cuprates, where it is confirmed that the Kohler's rule is satisfied inside the pseudogap phase of these compounds \cite{Chan}.

The Hall current, defined by $J_H^x=\sigma_{xy}E_y$, can also be computed microscopically near the Fermi surface of the model. As a result, its expression can be written as
$J_H^x=e\sum_{\mathbf{p}\sigma}\bar{\psi}_{\mathbf{p}\sigma}\mathcal{M}^{-1}_{xy}(\delta {p_y}-e{A_y})\psi_{\mathbf{p}\sigma}$,
where the quasiparticle velocity has been expanded in terms of the effective mass tensor $\mathcal{M}_{xy}^{-1}=(\partial^2 \bar{\varepsilon}_{\mathbf{k}}/\partial k_x \partial k_y)$. In this way, we can conclude that, when the Fermi surface of a given model possesses flat regions -- as it happens, e.g., in the so-called hot spot model \cite{Chubukov1,SSLee,Metlitski,Freire2} -- both the Hall current and the Hall conductivity $\sigma_{xy}$ will not have contributions from these specific regions.
This fact will be important for the considerations that follow in the present work.

A quantity of central interest in the present analysis is the so-called Hall angle $\theta_H$ that is
given by $\tan\theta_H=\sigma_{xy}/\sigma_{xx}$. In the case where there are no hot spots at the Fermi surface of the model, this naturally yields $\tan\theta_H=(B\chi^{-1}_{J_x P_x})/\rho_{xx}$.
As was already mentioned before, in the pseudogap phase of the cuprates, the quasiparticles at the hot spots are destroyed at low energies. Notwithstanding this, the corresponding dc resistivity will be given by $\rho_{xx}\sim T^2$. For this reason, the Hall angle will naturally become $\tan\theta_H\sim 1/T^2$ for this phase. 
These results are clearly consistent with the experimental data \cite{Ando,Raffy,Greven,Mirzaeia,Abdel-Jawad}.

We now move on to the most puzzling aspect of the transport properties of the strange metal phase observed in many correlated systems.
For this case, apart from already discussed subleading contributions, we propose that the Hall angle should be given by

\vspace{-0.3cm}

\begin{eqnarray}
\tan\theta_H\sim\frac{\rho^{hot}_{xy}}{\rho^{hot}_{xx}}+\frac{\rho^{disorder}_{xy}}{\rho^{disorder}_{xx}},
\end{eqnarray}

\noindent with the Hall resistivity $\rho_{xy}$ being naturally described by the expression $\rho_{xy}=[\sigma_{xy}/(\sigma_{xx}^2+\sigma_{xy}^2)]\approx \sigma_{xy}/\sigma_{xx}^2$. Therefore, we deduce from it that if there were only hot spots in the model, there would not be any contribution to the Hall angle, since the Hall resistivity $\rho_{xy}^{hot}$ would naturally vanish for this case. This happens
because of the emergent particle-hole symmetry at the low-energy fixed point in the model and the resulting dynamical nesting of the corresponding renormalized Fermi surface (for explanations about this change of topology of the underlying Fermi surface that was obtained by several RG approaches for the hot spot model, see the Refs. \cite{Chubukov1,Metlitski,SSLee,Freire2}).
This fact underscores the importance of the emergent charge conjugation $\hat{\mathcal{C}}$ symmetry at the hot spots in the low-energy effective theory.

By contrast, as weak disorder is included in the present model, there appears a temperature-independent contribution to the Hall resistivity $\rho_{xy}^{disorder}$. Hence, in view of the fact that this latter relaxation mechanism results in a dc resistivity given by $\rho_{xx}^{disorder}\sim T^2$, we
may conclude that the Hall angle at least at optimal doping would also become described by $\tan\theta_H\sim 1/T^2$ within this scenario. This could provide a rationale for the explanation of the thus far mysterious separation of lifetimes, which is observed experimentally in many correlated systems reported in the literature \cite{Ando,Raffy}. Note that the above result for the Hall angle depends crucially on the mechanism for the emergent (exact) particle-hole symmetry that exists in the low-energy fixed point of the hot spot model. Small deviations from the exact charge conjugation symmetry would imply that another contribution to the Hall angle would also appear. In other words, one would obtain in the most general way that $\tan \theta_H\sim c_1/T+c_2/T^2$ (for $c_2\gg c_1$ at optimal doping) inside the strange metal phase of the model. This picture seems consistent with many experimental observations, where a deviation from the quadratic $T$-dependence in the inverse Hall angle is clearly visible with further doping (see, e.g., Refs. \cite{Ando,Raffy,Ando2}).
In this respect, it is interesting to draw attention to the fact that the present scenario resonates with a recent calculation that employs nonperturbative holographic methods to model possible strange metals described by gravitational theories in a spacetime with one extra dimension, in which the role of charge conjugation symmetry for the computation of the Hall angle has been emphasized \cite{Blake}. We also mention the work by Coleman \emph{et al.} \cite{Coleman} that originally pointed out to the crucial role of the charge-conjugation symmetry for the correct interpretation of the apparent two relaxation times \cite{Anderson} that naturally emerge in quantum critical materials such as, e.g., the cuprate superconductors.

\textbf{Conclusions.} -- In this work, we have calculated the Hall angle and the magnetoresistance as a function of temperature and magnetic field for a spin-density-wave quantum critical theory
in the presence of weak disorder. We have used the memory-matrix scheme to evaluate all transport quantities, without assuming the existence of
quasiparticles at low energies in the model. As a result, we have applied this non-quasiparticle transport theory to discuss the important case of the anomalous metallic phases of the cuprate superconductors from the point of view of their fermiology.
We have argued that the Hall angle $\theta_H$ is given by $\tan \theta_H\sim 1/T^2$ inside the pseudogap phase, and $\tan \theta_H\sim c_1/T+c_2/T^2$ (for $c_2\gg c_1$ at optimal doping) in the strange metal phase. The proposed scenario is in qualitative agreement with experiments in many cuprate compounds.  As a next step, we plan to apply the present formalism also to the important case of the pnictide superconductors, whose multi-band character makes the present analysis a bit more involved. In this way, we expect that the present approach may provide a unified picture of SDW quantum criticality in many correlated systems. 

\acknowledgments

I would like to thank the Brazilian agency CNPq under grant No. 405584/2016-4 for financial support.

\end{document}